\documentclass[conference]{IEEEtran}
\IEEEoverridecommandlockouts

\usepackage{enumitem}

\usepackage{cite}
\usepackage{amsmath,amssymb,amsfonts}
\usepackage{algorithmic}
\usepackage{graphicx}
\usepackage{textcomp}
\usepackage{xcolor}
\usepackage{subcaption}
\def\BibTeX{{\rm B\kern-.05em{\sc i\kern-.025em b}\kern-.08em
    T\kern-.1667em\lower.7ex\hbox{E}\kern-.125emX}}
\begin{document}

\title{Multi-Scale Asset Distribution Model for Dynamic Environments 
}

\author{\IEEEauthorblockN{Payam Zahadat}
\IEEEauthorblockA{\textit{Computer Science Department} \\
\textit{IT University of Copenhagen}\\
Copenhagen, Denmark \\
paza@itu.dk}
\and
\IEEEauthorblockN{Ada Diaconescu}
\IEEEauthorblockA{\textit{Telecom Paris, LTCI} \\
\textit{Institute Polytechnique de Paris}\\
Paris, France \\
ada.diaconescu@telecom-paris.fr}
}

\maketitle

\begin{abstract}
In many self-organising systems the ability to extract necessary resources from the external environment is essential to the system's growth and survival. Examples include the extraction of sunlight and nutrients in organic plants, of monetary income in business organisations and of mobile robots in swarm intelligence actions. When operating within competitive, ever-changing environments, such systems must distribute their internal assets wisely so as to improve and adapt their ability to extract available resources. 
As the system size increases, the asset-distribution process often gets organised around a multi-scale control topology. This topology may be static (fixed) or dynamic (enabling growth and structural adaptation) depending on the system's internal constraints and adaptive mechanisms.
In this paper, we expand on a plant-inspired asset-distribution model and introduce a more general multi-scale model applicable across a wider range of natural and artificial system domains. 
We study the impact that the topology of the multi-scale control process has upon the system's ability to self-adapt asset distribution when resource availability changes within the environment. 
Results show how different topological characteristics and different competition levels between system branches impact overall system profitability, adaptation delays and disturbances when environmental changes occur.     
These findings provide a basis for system designers to select the most suitable topology and configuration for their particular application and execution environment.

\end{abstract}


\section{Introduction}

Autonomic and self-organising systems  
must manage available resources from their environment 
and invest them efficiently into internal assets,  
to ensure their growth, competitiveness and survival~\cite{bejan2012}. Examples range from natural systems such as trees, through social systems such as business organisations, and all the way to cyber-physical systems (CPS) such as collaborative robot swarms.  To improve their resource intake within dynamically changing environments, such systems must often self-adapt their internal structures and their \textit{asset distribution} within those structures. To ensure viability as the amount of managed  assets increases, the system's self-adaptive control often takes the form of a multi-scale topology.

Multi-scale structures (e.g. \cite{saso2019,fgcs2021}) include multiple abstraction levels, where each level increases the granularity of observation of the level below -- i.e. information about the lower scale (micro) is lost in the abstraction process to the higher scale (macro). This allows multi-scale systems to increase their scopes, or operation domains, while limiting the amount of resources needed to handle information at each scale. 
Hence, multi-scale schemes can be applied to achieve system-wide coordination among numerous self-adaptive processes in large systems  \cite{fgcs2021} (e.g. adaptive asset distribution here). 

While numerous self-adaptive algorithms exist for context-aware asset distribution in various application domains (sec. \ref{sec:rel-work}), much less is known about the various impacts that the \textit{topology} of the multi-scale control system has on the self-adaptation process in general. This paper aims to provide the basis for such analysis by identifying several key topological features and linking them to generic characteristics of the self-adaptation process (e.g. reactivity and costs). 

\begin{figure}[bp]
\includegraphics[width=0.49\textwidth, angle=0]{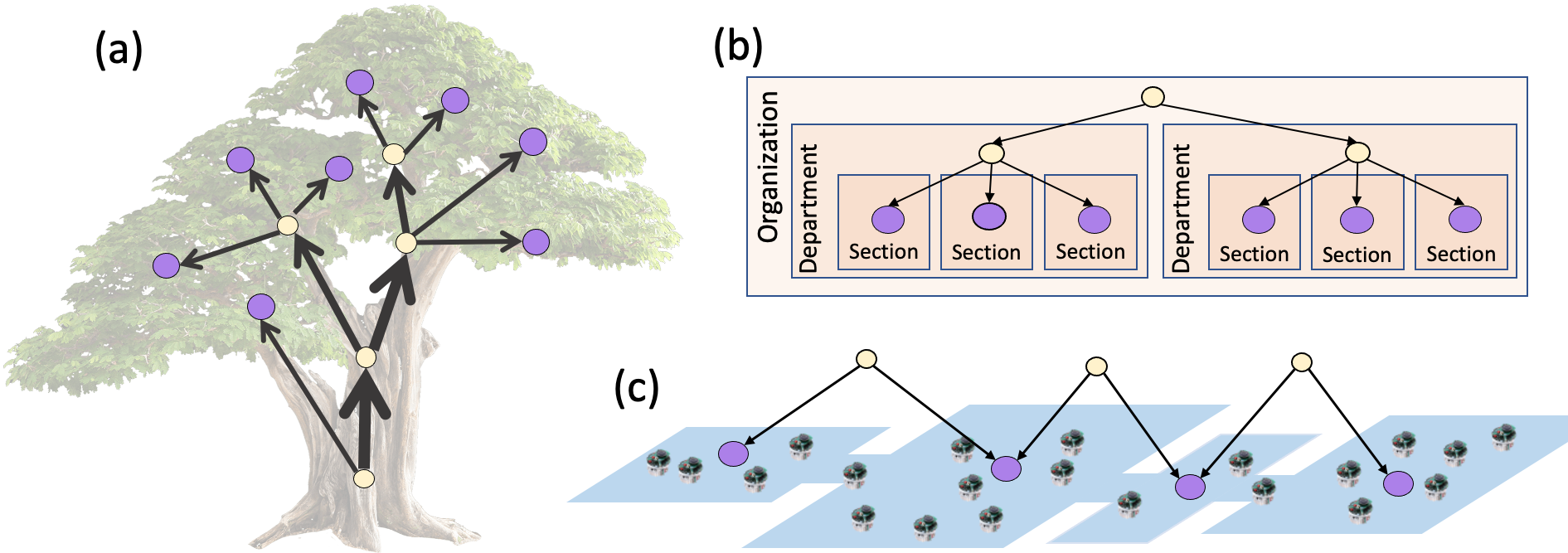}
\caption{Examples of systems that feature an adaptive distribution of internal assets. a) adaptive nutrient flows within a tree; b) adaptive budget investments within a business organisation; c) adaptive distribution of robots across various regions. 
}
\label{fig:examplesystms}
\end{figure}

We consider three illustrative examples (Fig.\ref{fig:examplesystms}) of systems featuring self-adaptive asset distribution, to guide an initial selection of topologies to analyse. 
First, we 
consider a common example from natural systems \cite{zahadat2017gecco}: trees absorb sunlight, water and mineral resources and transform them into organic matter, in turn forming internal structures that are essential to growth and survival. Hence, depending on resource availability in the environment, trees self-adapt their growth process to prioritise development towards resource-rich areas (e.g. leafy branches growing towards sunny patches and roots towards moist, mineral-rich soils). 
Similarly, within the socio-economic realm, business organisations must employ their assets (e.g. workers) to fulfill service requests from their market environment, so as to grow and survive within a competitive context \cite{ALife2020}. They must self-adapt to unexpected fluctuations in market demands by reallocating assets to the most popular service sectors.  
Finally, we consider a cyber-physical systems (CPS) example: a robot swarm must coordinate their actions across several rooms to achieve a shared task (e.g. cleaning). Robot distribution across the rooms should self-adapt to the level of cleaning services required within each room, which may dynamically change~\cite{zahadat2013ECALPSI,zahadat2016adaptivebehavior}.

In previous works \cite{zahadat2019SASO,ALife2020}, we proposed a decentralised asset distribution algorithm for systems featuring tree-like multi-scale control structures (e.g. tree plants and business organisations). Here, the system's internal assets 
were employed to maintain and to grow system branches. Feedback from the branches indicated their efficacy in acquiring external resources. This feedback was used to skew asset distribution towards more successful branches. e.g. tree branches getting more sun received more nutrients for further growth. Business departments 
making more profits received more workers for reinforcing their capacity.

This paper capitalises on this initial experience to generalise our asset distribution algorithm for more diverse structures. This allows its applicability across a wider range of topologies and self-adaptive features. Based on the new algorithm, we aim to study the impact of system topology on the context-aware self-adaptation of asset distribution. 
We consider a control system's topological variation along three main dimensions: 
\begin{itemize}
    \item Single or multiple roots: resulting in tree topologies or directed acyclic graphs, respectively;
    \item Static or growing topology: in static cases, assets are merely relocated within a pre-existing system structure; whereas in growing cases, assets also contribute to structural self-adaptation. 
    \item  Single or multiple control scales: resulting in single or multiple decision levels.
     
\end{itemize}
From the combinatorial space of these dimensions, we select several concrete topologies that we consider representative for natural and artificial systems (Cf. illustration in Fig. \ref{fig:topologies}). On the one hand, we analyse several tree topologies for plants and business organisations -- i.e. single-root with growing multi-scales, with static single-scale, or with static multi-scales. On the other hand, we consider several multi-root topologies with a single control scale, as representative for robot swarms carrying-out collective tasks -- i.e. `linear', `circular' or `complete' topologies (Fig. \ref{fig:topologies}). Here, control coordination between roots only occurs indirectly via observable impacts on the shared entities they control. This corresponds to the fact that robots in different rooms only coordinate via individual representatives located at the doors between rooms.

\begin{figure}[!b]
  \centering
    \includegraphics[width=0.4\textwidth, angle=0]{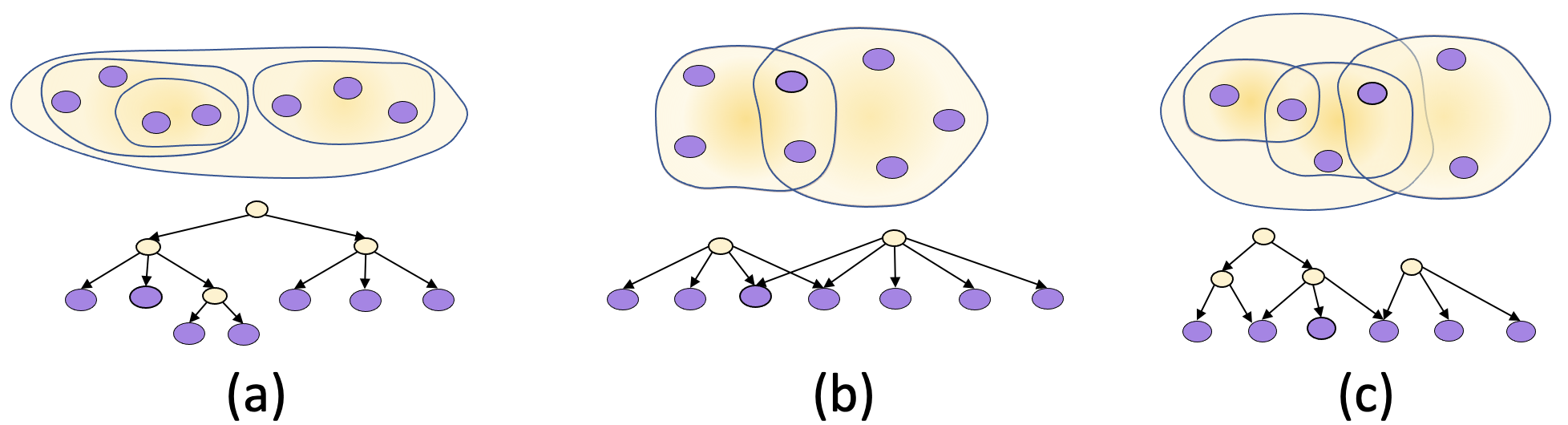} 
\caption{Examples of various systems' topological structures}
\label{fig:systemtypes}
\end{figure}

We study the self-adaptability of these topologies to variations in resource availability within the environment. 
We focus on environment changes that require system-wide coordination between self-adaptation processes (rather than being addressable via local self-adaptation). This choice allows studying the impact of the entire system control topology (rather than just local control sub-trees). 
In all cases, we introduce 
self-adaptation delays within controlled entities to emphasise the role of the control topology in the self-adaptation behaviour.

Results show the impact of system topology on key self-adaptation features, notably including: the \textit{extent} of self-adaptability to context changes; the \textit{duration} of self-adaptation; and the \textit{disturbance in efficiency} during self-adaptation. Some of the key findings include (Cf. sec. \ref{sec:discuss}): 
\begin{enumerate}
    \item Growing topologies self-adapt better to new opportunities (via asset re-specialisation) and provide more profits than static ones. This comes at the price of longer adaptation delays and profit disturbances during adaptation.  
    \item  Multi-root topologies achieve global control coordination slower than tree topologies. This incurs longer adaptation delays and more profit disturbance.  Similar effects would be observed in multi-scale topologies when including inter-scale communication delays (e.g. studied in \cite{acsos2021}). 
    \item While some level of performance-oriented competition is necessary to drive self-adaptation, too much competition actually hampers self-adaptation by creating too much inertia and preventing the detection of new opportunities.
\end{enumerate}

These findings provide a basis for system designers to reflect upon and select the most suitable topology for their particular application and execution environment, considering all the advantages and limitations of each topological choice. They also provide opportunities for further studies on larger and more complex topologies (e.g. scale-free or community networks), as found in many real-world systems (e.g. \cite{barabasi2003}).

\section{Related Works}
\label{sec:rel-work}

Asset distribution is a broad research topic covering numerous application domains (e.g. from dynamic VM allocation in cloud systems \cite{kounevSEAMS2011} and process scheduling in mixed-criticality real-time systems \cite{pautet2018}; through adaptive data-mediation \cite{cube-saso2012} and power networks \cite{BALE2015}; and all the way to group organisation in social insects\cite{Julian99} or swarm robotics \cite{zahadat2016adaptivebehavior}).  We can only mention a few of these approaches here, emphasising some of their key characteristics relevant to our study. Still, to the best of our knowledge, only limited studies assess the importance of the control topology in asset distribution approaches that are self-adaptive and  multi-scale. e.g. \cite{arlesJAAMAS21} studies the impact of topology on decentralised data-collection 
in complex networks, yet these are non-adaptive and single-scale.  

Concerning self-adaptive resource allocation in static (non-growing) structures,
\cite{diStefano2005} proposes decentralised resource allocations in grid networks based on a spatial algorithm optimisation approach. \cite{energyDistrib2015} targets power grids  to optimally distribute produced electricity to users, minimising their costs while covering their demands. \cite{kounevSEAMS2011} proposes self-adaptive resource-allocation in virtual environments, to deal with dynamically deployed services and their workload fluctuations by adding/removing application servers to clusters and virtual CPU cores to virtual machines (VMs). 
\cite{sardes2006} also deals with resource allocation in clustered servers, by self-adapting the number of replicated databases in a clustered J2EE application when the load varies.

Resource distribution solutions with self-adaptive structures (e.g. growing), may concern, for instance, the long-term extension and restructuring of energy system infrastructures, by investing into new power plants to match shifting demands \cite{BALE2015}.  In data-mediation systems, \cite{cube-saso2012} investigate self-growing and self-adapting structures as means to address unexpected changes in data sources/consumers, workloads and servers.

Another category of resource allocation solutions concerns self-adaptive group formation from members of social organisations that aim to achieve a collective task. This involves the self-adaptive distribution of members into different task groups in, e.g. social insects (honeybees \cite{Seeley82,Huang92}, wasps \cite{Torres2012}, termites \cite{Crosland2010}, and ants \cite{Julian99}). Similarly, in technical systems inspired by social insects, swarm robots self-coordinate to dynamically allocate themselves to various regions depending on the dynamic demands imposed internally by the swarm or externally by the environment \cite{JonesM03,Yang2009,zahadat2013ECALPSI,zahadat2016adaptivebehavior}.

In all relevant cases, distributed assets are conserved (energy, money, physical resources), whereas control information may not be (because of various multi-scale abstractions).  

\section{Background and Previous Work}

\subsection{Multi-scale control systems}
\label{subsec:multi-scale}

Multi-scale structures (e.g. \cite{saso2019,fgcs2021}) include multiple abstraction levels, where each level increases the granularity of observation of the level below (i.e. coarse graining). As higher scales often observe larger domains, or scopes, than lower scales, inter-scale  abstraction allows them to avoid an ever-growing need for information storage and processing resources. 
e.g. a world map provides fewer details than a city map, so that both maps can fit onto an A3 page. Hence, the multi-scale principle allows observation systems to scale-up with the size of their observation domain.  

\begin{figure}[tbp]
\centerline{
\includegraphics[width=0.4\textwidth, angle=0]{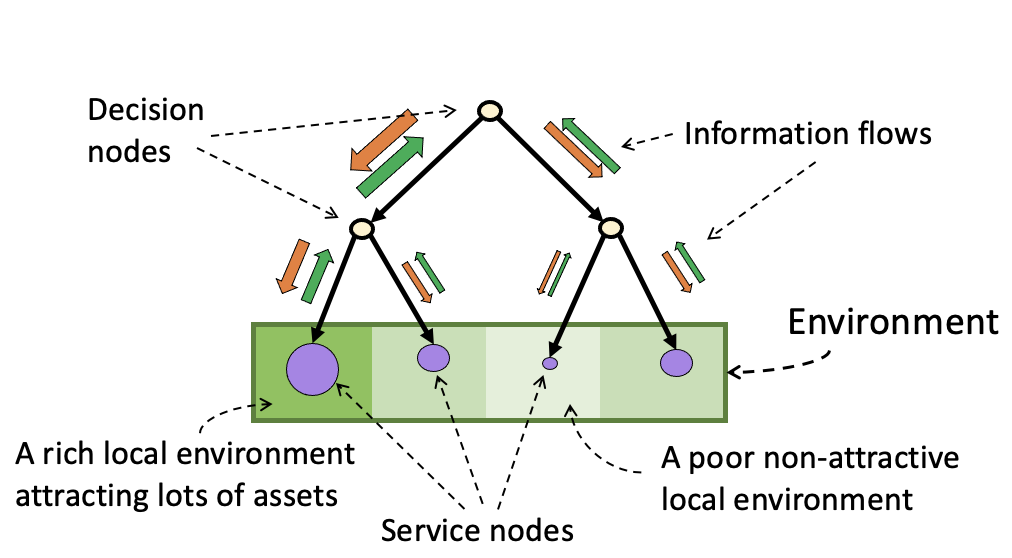}
}
\caption{A multi-scale system distributing internal assets in a way that is well-adapted to its environment.}
\label{fig:graphmodel}
\end{figure}

This structural principle also applies to multi-scale \textit{control} systems, which are characterised by two main information flows \cite{fgcs2021}: i) \textit{observation abstraction flow} (bottom-up), as above -- where an entity at a higher-scale (macro) collects and abstracts information about several entities at the lower scale (micro); and, ii) \textit{control flow} (top-down) -- where a higher entity (macro) provides self-adaptation directives to lower entities (micro). Directives rely on decisions taken based on abstracted information from their micro entities (bottom-up observation flow) and adaptation directives from their macro entities (top-down control flow). 
Hence, the two flows -- abstraction and control -- form multiple control feedback-loops between subsequent system scales (e.g. Fig.\ref{fig:graphmodel}). 

Such multi-scale control schemes enable system-wide coordination among self-adaptive processes while limiting the amount of resources required at each decision node (e.g. \cite{fgcs2021}). This explains their wide-spread occurrence in complex self-adaptive systems, including the large-scale asset distribution systems that we study here.

\subsection{Distribution of internal assets}

A system contains various internal assets,
which enable its functions and ensure its survival.
The manner in which these assets are distributed among different system parts is essential to the system's development, adaptability and survival in the face of changing conditions. The most effective investment of available assets within various system parts depends on the system's internal structure (e.g. topology) and its external environment (e.g. available resource distribution). 
To illustrate these ideas, we consider three examples -- from the natural,  business, and artificial world (Fig.~\ref{fig:examplesystms}) -- and use an analogy between them to highlight the essential roles and behaviours of internal asset distribution.

The natural example consists of 
a plant (e.g. a tree), where 
water and minerals provided at the root represent internal resources, or assets. These are distributed throughout the plant and `invested' in various branches to produce sugars (for energy provision) and hence to enable growth and further branching. 
The way in which available assets are allocated among various branches depends on the local 
state and context 
of each branch -- e.g. local sunlight availability, meaning that the more access to sunlight a branch has the better it grows. 

A similar phenomenon can be observed within a business organisation.  
Here, internal assets consist of workers and the budget available to hire them. The budget is allocated to hire and remunerate workers for each service that the organisation offers. It is distributed among these services depending on the demand and profitability of each service.

The final example is a collective of robots that are assets for taking care of given tasks in different regions of an arena. The distribution of the robots among these different regions depends on the amount of work needed in each region.

The abstract model introduced in this study is a generalized version of a plant-inspired model
that is introduced previously\cite{zahadat2017gecco} and summarized in the following sub-section.

\subsection{Plant-inspired model of asset distribution}
\label{sec:VMC}

The overall shape of a plant results from a combination of genetically-encoded features, physical and chemical mechanisms, and environmental features, including available resources \cite{zahadat2017gecco}.
A plant has a root and a tree-like structure of branches above the ground. 
The root provides the aerial plant with 
essential nutrients 
i.e. water and minerals.
The nutrients are transported via vessels within the plant's branches. 
The nutrients are the plant's assets to be distributed internally and favour differential branch growth.  
The leaves on each branch use sunlight to convert these assets into sugars (providing the plant with energy) -- via the process of photosynthesis. 
The plant's growth process relies not only on the assets (i.e. nutrients) reaching the leaves of each branch, but also on the local environmental resources (e.g. sunlight) available at the branch.
The growth process shapes its crown so as to improve its leaves' access to sunlight. 
The plant's shape can be abstracted as a directed acyclic graph (Fig.~\ref{fig:examplesystms}a): branches, including the vessels inside them, correspond to graph edges; the plant's tips to the graph's leaf nodes; and the branch junctions to the graph's non-leaf nodes.

The growth process relies on two 
flows with opposite directions: 
1) a forward-flow through the edges (plant vessels), from the root towards the leaves (plant tips), carrying common assets (nutrients); and 
2) a backward-flow from the leaves (plant tips) towards the root carrying information about the attractiveness of the sub-trees, i.e. access of each sub-tree to local resources (sunlight). %
These flows correspond to the two multi-scale flows (subsec. \ref{subsec:multi-scale}), where (1) asset distribution from roots to leaves  belong to the control flow that adapts the tree shape, based on previous observation flows (2) from leaves to the root informing about sub-tree attractiveness.  

In real plants, the backward information flow is realized via a hormone produced at the tips based on the intensity of sunlight. It flows along the plant's vessels and incrementally changes their thickness  -- more hormone flow leads to a thicker vessel over time. 
This mechanism allows junctions to estimate the attractiveness of each branch, informing the future investment of the assets in that part of the plant.
At the branch junctions, the differences between the vessel thickness of the various branches determine the distribution of the nutrients among them -- thicker vessels take a larger share of the assets available at that junction. 

In the plant-inspired model \cite{zahadat2019SASO,ALife2020}, the vessel thickness of every branch is abstracted as a variable that is attributed to the corresponding edge. This variable changes incrementally in response to the values of the backward flow passing through that edge. 
The non-leaf nodes, or junctions, 
decide on the distribution of assets among their children. They do this by considering the corresponding variables of their outgoing edges to their children. 
Therefore, we refer to these nodes as \emph{decision nodes} (see Fig.~\ref{fig:graphmodel}). 

The leaf nodes are the main points of interaction of the system  with the local environment. The backward-flows are initiated at each leaf node as a function of the assets invested at the node and the quality of the local environment. Investing more assets at the leaves that are located within more interesting regions (i.e. richer in resources) of the environment benefits the system.
As leaf nodes are the production sites of the backward-flows, which indicate the investment profitability within each corresponding region, 
the leaf nodes are also called \emph{service nodes} (Fig.~\ref{fig:graphmodel}).  

The system's growth model is designed so that the corresponding system graph only grows at the leaf nodes and only if the invested assets at the leaf node crosses a given threshold. 
Likewise, the system can shrink by removing all the leaves of a node, if the total assets invested in 
these child leaves is below a given threshold. The node will then become a leaf node holding all the assets previously invested in its children.

\subsection{Various types of system structures}

Many complex systems, such as plants and large-scale business organisations, feature some form of tree-like 
structure.    
Above, we have already highlighted the tree topology of plants.
Similarly, large business organisations are often also structured into tree-like topologies -- divided into departments, with each department further divided into several sections, in turn divided into sub-sections, and so on (Fig.~\ref{fig:examplesystms}b). Such tree structures are a particular kind of multi-scale control topology.

On the other hand, some systems do not have a tree-like structure. An example is a group of robots (Fig.~\ref{fig:examplesystms}c) distributed across a number of rooms to perform some collective tasks, e.g. keeping the rooms clean. 
Once there is a lack of tasks
in a room (e.g. the room is already relatively clean), the robots may start moving to the neighboring rooms where there are more tasks to perform (e.g. dirtier rooms). 
The decision for robot redistribution can be taken by an exogenous agent located at the gates between neighboring rooms, or it can happen via direct interactions between robots that  
exchange information through the gates. 
In either case, the decision-making process based on the information from  the neighboring rooms can be abstracted and represented as a decision node. 

In the example in Fig.~\ref{fig:examplesystms}c, decision nodes are the roots of the graph. The system has multiple roots, meaning that there is no higher level of coordination amongst those decision nodes. 

When systems need to adapt to internal and external changes -- e.g. internal growth, external resource fluctuations -- their 
structure may change accordingly, 
e.g. similarly to the shaping of a plant crown.
For example, an organisation may subdivide one of its sections into subsections if the number of workers in that section becomes too large to administer by a single manager; or in case further specialisation of experts is needed.

Fig.~\ref{fig:systemtypes} exemplifies several systems with various 
types of control topology. 
Fig.~\ref{fig:systemtypes}a shows a tree-like structure with a single root.  
Fig.~\ref{fig:systemtypes}b Shows a multi-root system with a single scale of decision making (a.k.a control). 
Fig.~\ref{fig:systemtypes}c depicts a rather complex multi-scale system with multiple roots.
In addition to the graph representation, Fig.~\ref{fig:systemtypes} also depicts the systems as overlapping decision sets, to give an alternative view of the interactions between the service nodes.
Every decision set is equivalent to a sub-tree (including a parent node) in the graph representation.

\section{Multi-Scale Asset Distribution model}

The model we introduce here, namely, the Multi-Scale Asset Distribution (MSAD) model, is more general than the plant-inspired model presented in sec.~\ref{sec:VMC}. 
Similar to the plant model, we represent a system generically as a directed acyclic graph that (potentially) can grow. 
The decisions on asset distribution are taken at the \emph{decision nodes} based on information flows arriving via the graph edges, in a bottom-up and top-down manner. 
The \emph{service nodes} are the lowest-level entities of the system located at the interaction points of the system with the environment. 

Unlike the previous model, assets here are modeled explicitly. That is, there is a distinction between the forward-flow that carries information about the assets, and the actual assets which can move within the system according to local decisions. Additionally, the backward-flow carries two pieces of information: one related to the attractiveness of the sub-trees (i.e. profitability); and the other related to the estimation of the assets already invested at the sub-trees.  
The definition of assets, profitability, and flows, as well as the mechanisms involved in the redistribution of assets are detailed below.

\subsection{Assets}
Assets in the MSAD model are equivalent to the system's workforce, either directly (e.g. as workers) or indirectly (e.g. as budget used for hiring workers). 
In decision nodes, assets are mainly used for performing management tasks. In service nodes, assets are mainly used to provide services that produce some form of profit for the system.
In general, assets can be added or removed, but once in the system, they follow the principle of mass conservation. 

\subsection{Environment and profitability} 

Service nodes are the main interaction points between system assets and the external environment, where interactions acquire profit for the system. 
The local environment of each service node 
is considered as a local resource -- i.e. more local resource 
at a service node leads to more profit for the same amount of assets. 
In the plant example (Fig.~\ref{fig:examplesystms}a), more
sunlight reaching a branch increases the conversion of nutrients (assets) into sugars within that branch\footnote{in plants, assets are constantly added to the system from the roots and turned into sugars at the leaves.}. This leads to further growth and branching, and thus to more hormones production, indicating the higher profitability of that region.
In the business organisation example (Fig.~\ref{fig:examplesystms}b), 
the budget (asset) is used at the service nodes to hire workers that provide services to customers and hence acquire profits for the organisation.   
In the example of robots cooperating across several rooms 
(Fig.~\ref{fig:examplesystms}c), the robots (assets) operate as groups, where each group (service nodes) performs requested tasks within one room (local environment). The amount of work done within each region indicates its profitability.

\subsection{Information flows}

\begin{figure}[htbp]
\centerline{
\includegraphics[width=0.16\textwidth, angle=0]{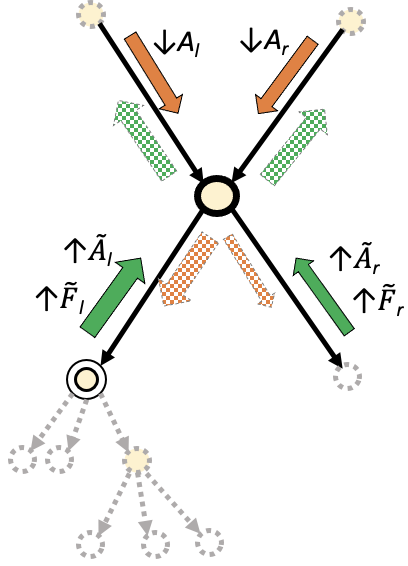}
}
\caption{An example decision node (thick circle) with incoming and outgoing flows. 
The bottom-up and top-down flows are respectively depicted in green and orange. The entering and exiting flows are respectively in solid and hatched colors. 
}
\label{fig:flows}
\end{figure}

The proposed model 
relies 
on three information flows (Fig. \ref{fig:flows}). 
The direction of each information flow is indicated via the name of its corresponding variable as: an $\uparrow$ sign for bottom-up flows and a $\downarrow$ sign for top-down flows. 
The three flows are:
\begin{enumerate} 
    \item $\uparrow\!\tilde A$: estimation of existing assets
    \item $\uparrow\!\tilde F$: estimation of profitability
    \item $\downarrow\!A$: amount of eligible assets  
\end{enumerate}

Fig.~\ref{fig:flows} exemplifies a decision node with two parents and two children.  
The values of the flow variables $\uparrow\!\tilde A_l$ and $\uparrow\!\tilde F_l$ represent the estimation of the contained assets and of the profitability in the left sub-tree.  
The flows $\uparrow\!\tilde A_r$ and $\uparrow\!\tilde F_r$  
represent the equivalent variables in the right sub-tree. 
Variable $\downarrow\!A_l$ represents the amount of assets that the node is eligible to hold, according to its left parent. 
The $\downarrow\!A_r$ is the equivalent variable from the right parent. 

In practice, updating the outgoing flows is often subject to time delays. 
The delayed information update can occur either suddenly after a delay period (hard delay), or gradually over a period of time (soft delay) -- Cf. details below.
In the following, the computation of each flow is described in more details.
The system dynamics are implemented via discrete time steps.

\subsubsection{$\uparrow\!\tilde A$, estimation of assets existing within a sub-tree}
This information flow starts at the service nodes (leaves) and flows toward the root(s). 
The flow from each node to its parent(s) is an estimation of the total assets available at that node and all of its children -- i.e. the whole sub-tree of that node. 
In every decision node, at every simulation time $t$, the value of the output flow is updated. The value is updated such that it gradually approaches a target value,
$AA(t)+\sum_{i\in \text{children}}{\uparrow\!\tilde A_i(t)}$,
which is the total sum of the assets currently located at the decision node ($AA$), and the current values of the incoming flows from all the children of that node. 
To implement the gradual approach to the target value, the difference ($\Delta_A(t)$) between the target value and the current value of the output flow ($\uparrow\!\tilde A_{out}(t-1)$) is computed as:  
\begin{equation}
    \Delta_A(t) = AA(t)+\sum_{i\in \text{children}}{\uparrow\!\tilde A_i(t)} - \uparrow\!\tilde A_{out}(t-1)
\end{equation}

$\Delta_A(t)$ is then used to compute the new value of the output: 
\begin{equation}
    \uparrow\!\tilde A_{out}(t) := \uparrow\!\tilde A_{out}(t-1) + \gamma_{\uparrow\!A} \cdot \Delta_A(t)\\
    \label{eq:updateA_estimated}
\end{equation}
where $\gamma_{\uparrow\!A}\in[0,1]$, is a weighting factor determining the rate of change toward the target value. The weighting factor is responsible for a soft delay in information communication between nodes and their parents.
A hard delay can be included by delaying the update of the output flows for a given period (not implemented here).

\subsubsection{$\uparrow\!
\tilde
F$, estimation of profitability within a sub-tree}
This information flow behaves similarly to $\uparrow\!\tilde A$, starting at the service nodes and flowing toward the root(s). 
The output flow from each node to its parent(s) is: an estimation of the profitability of the node, if it's a service node; and an estimation of profitability of all the children of the node, if it's a decision node. 
In every decision node, at every simulation time $t$, the value of the output flow is updated to gradually approach a target value $\sum_{i\in \text{children}}{\uparrow\!\tilde F_i}(t)$.
Similar to $\uparrow\!\tilde A$, the  difference between the target value and the current value of the output flow is computed ($\Delta_F(t)$) and is used  to compute the new value of the output flow, as:

\begin{equation}
\uparrow\!\tilde F_{out}(t) := \uparrow\!\tilde F_{out}(t-1) + \gamma_{\uparrow\!F} \cdot \Delta_F(t)\\
\end{equation}
\begin{equation*}
\Delta_F(t) = \sum_{i\in \text{children}}{\uparrow\!\tilde F_i}(t) - \uparrow\!\tilde F_{out}(t-1)
\end{equation*}
where $\gamma_{\uparrow\!F}\in [0,1]$ is a weighting factor inducing a soft delay in information communication. 

\subsubsection{$\downarrow\!A$, eligible amount of asset, according to a parent}
This information flow reflects 
local system decisions to control the asset distribution. 
It starts at the roots based on the estimation of total assets within their sub-trees ($\uparrow\!\tilde A$). At every decision node, the total flow arriving from all the parents of the node is summed up ($\sum{\downarrow\!A}$) and its value is corrected based on the estimation of the total assets within the sub-tree of the node:
\begin{equation}
    \downarrow\!A':= \min{(\sum_{i\in \text{parents}}{\downarrow\!A_i} \text{ , } \sum_{j\in \text{children}}{\uparrow\!\tilde A_j+AA)}}
\end{equation}
where $\downarrow\!A'$ is the corrected value, 
$AA$ is the amount of assets currently located in this node. 

Then, the amount of assets that is needed at the node (e.g. management costs, $c$) is 
subtracted  
and the remaining flow is divided between the child nodes based on the estimated profitability ($\uparrow\!F$) of each child and on a competition factor:
\begin{equation}
\downarrow\!A_k'' := \max{(0,\downarrow\!A' - c) \cdot \frac{{\uparrow\!\tilde F_k}^\beta}{\sum_{j \in \text{children}}{{\uparrow\!\tilde F_j}^\beta}}}
\end{equation}
where $\downarrow\!A_k''$ is the amount of asset for the sub-tree of child $k$ computed based on the most recent information, $c$ is the node's asset consumption 
and $\beta$ is a competition factor. By increasing $\beta$ to values larger than 1 the competition between sibling nodes is intensified. $\beta=0$ means that all  siblings get the same amount of $\downarrow\!A$ flow irrespective of their profitability.

To take adaptation delay into account, 
the new value of $\downarrow\! A_{k}$ can be updated incrementally toward the value computed based on the latest information, $\downarrow\!A_k''$, as follows:

\begin{equation}
\downarrow\! A_{k}(t) := \downarrow\! A_{k}(t-1) + \gamma_{\downarrow\!A} \cdot \Delta\\
\end{equation}
\begin{equation*}
\Delta = \downarrow\!A_k'' - \downarrow\! A_{k}(t-1)
\end{equation*}

where $\Delta$ is the difference between the current and the target value, 
$\gamma_{\downarrow\!A}$ is a weighting factor inducing communication delay.

\subsection{Asset relocation}
The asset distribution process self-adapts to changes in the external environment and the internal conditions.  
Following the local decisions in the system, reflected by the top-down flow $\downarrow\!A$, assets may 
move between sibling nodes or between children and parents.
In the following, we describe the implementation of asset relocation operations used in this study. 

In the current implementation, assets 
are in one of two states: resident or non-settled. 
Resident assets, namely $AA^{resident}$, are bound to the node they are located at and are not ready to relocate. The non-settled assets, namely $AA^{nonset}$, are not settled and are in the process of relocation at a decision node.
As decision nodes abstract all 
information related to asset distribution among their children, in this implementation the asset relocation between sibling nodes is done only via their parents. 
The parent aggregates non-settled assets 
from its children 
and redistributes them based on the needs of the child nodes (Cf. details will follow). Extra assets will remain at the parent node and can be used later for its children or moved by the node's parents if needed.

The relocation process is implemented starting with the computation of a value, $\Delta P$, namely pressure difference, between every child and each of its parents. 
It is computed with respect to each parent as:
\begin{equation}
\Delta P = (\uparrow\!\tilde A - AA^{nonset}) - \downarrow\!A
\end{equation}
This value represents an estimated difference between the available and the expected (eligible) amount of assets at the node, indicating if (and how much) assets need to be exchanged between the parent and the child.
In this equation, $AA^{nonset}$ is decremented from $\uparrow\!\tilde A$ because $AA^{nonset}$ shows the amount of assets already in the process of relocation.

$\Delta P > 0$ means that there is a pressure to hand-over some assets to the parents, so as to reduce the difference. 
If the node is a service node, it will change the state of a fraction of its resident assets to non-settled and hand them over to the corresponding parent. The fraction equals to $\alpha \cdot \Delta P$ where $\alpha \le 1$ is the release factor of service nodes, leading to a soft delay in the relocation operation.
If the node is a decision node, it can hand over all the unused assets ($AA^{nonset}$). 

$\Delta P < 0$ indicates that the 
node needs to receive assets from its parents. 
After all children with a $\Delta P > 0$  make their attempt to reduce the pressure difference by transferring assets to their parent, the parent goes through all the children with $\Delta P < 0$ and moves the collected non-settled assets to them according to their need. 
The children receiving the non-settled assets may change them to resident, or keep them available for relocation to their children, depending on the needs.

It is not always possible to release all the pressure difference $\Delta P$ between children and parents. 
Likewise, some of the children may still need assets from their parents after the relocation. Such imbalances may be solved over time (in the next simulation steps).
Although the relocation operation can be performed within a time-scale that differs from the time-scale of the information flow updates, in the current study, an identical time-scale is used for both operations.

\subsection{Order of operations}
In the current implementation, at every simulation step $t$, the operations updating the information flows and relocating the assets are executed in the following order:
\begin{enumerate}
    \item 
    Update bottom-up information flows ($\uparrow\!\tilde A$ and $\uparrow\!\tilde F$) for all the nodes, starting from the highest depth toward the lowest (bottom-up order).
    \item
    Update top-down information flows ($\downarrow\!\tilde A$) for all the nodes, starting from the lowest depth (roots) toward the highest (top-down order).
    \item 
    Execute asset relocation operations at all the decision nodes, starting from the highest depth toward the lowest (bottom-up order).
\end{enumerate}

\section{Experimental setup}

\subsection{Environment}
\begin{figure}[t]
\centerline{
\includegraphics[width=0.4\textwidth, angle=0]{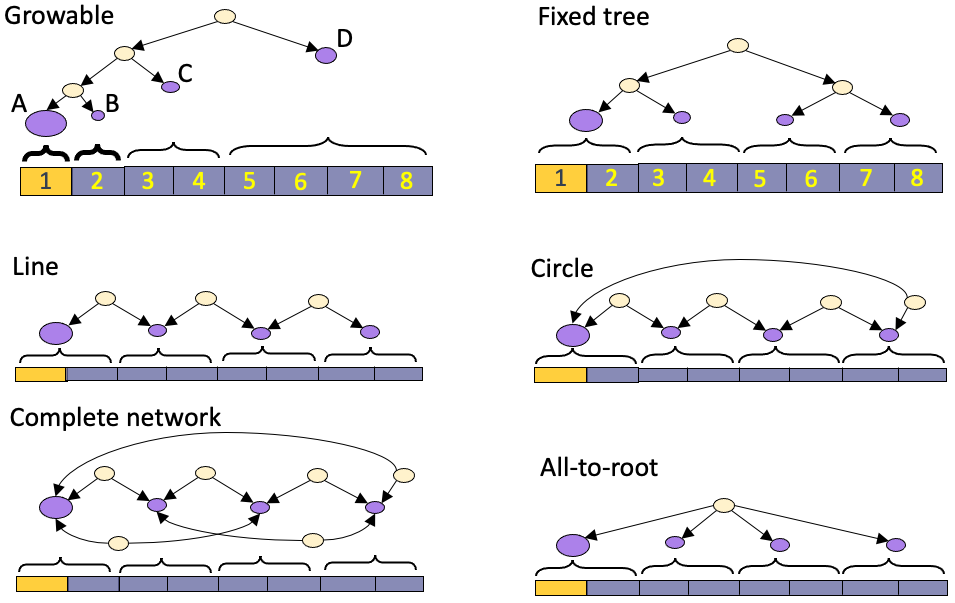}
}
\caption{Illustration of topologies used in the experiments.}
\label{fig:topologies}
\end{figure}

An environment is implemented as a 1-dimensional sequence of regions, $m=1..M$. 
In Fig.~\ref{fig:topologies}, the bars below each graph represent a sample environment with $M=8$ regions, numbered $1..8$. 
Every region $m$ has a quality value, $q_m$, that is the profitability of the region. 
This value indicates how much profit can be produced per simulation step if one unit of asset is fully invested in that region. 
The profitability of the regions are color-coded in Fig.~\ref{fig:topologies}, with yellower colors indicating higher profitability for the region. 

A service node can support services demanded in a single or multiple regions. 
In Fig.~\ref{fig:topologies}, the regions that are supported by each service node are indicated using curly brackets. In the example of the `fixed tree' topology, 
every service node supports two regions. 
On the other hand, in the example of `growable' topology, the service nodes A and B support only one region each; service node C supports 2 regions; and service node D supports 4 regions.
The profitability of a service node is computed as the product of its assets and the average profitability of the regions it supports. 
\begin{equation}
\tilde F_i = AA_i \cdot \frac{1}{|C|}\sum_{c\in C}{q_c}
\label{eq:profit}
\end{equation}
where $AA_i$ is the assets invested at service node $i$; $C$ is the set of regions supported by the service node; and $|C|$ is the number of such regions. 
The profit produced by the assets in a service node with multiple supporting regions is the average of the profit that can be produced in each of those regions separately (Eq. \ref{eq:profit}). 
This is
equivalent to a worker equally dividing their time between different services. 

In the examples of Fig.~\ref{fig:topologies}, the `growable' topology benefits from the further branching into service nodes A and B that specialise to correspondingly support regions 1 and 2. 
This branching allows larger amounts of assets to be invested specifically in the more profitable region 1.

The bars on top of Fig.~\ref{fig:examplegrowth} show the environments used in all the experiments reported in this paper.
Initially, the leftmost region of the environment has a higher profitability $q_1=0.3$ and all the other regions have a lower profitability $q_i=0.1, \forall i=2..8$ (Fig.~\ref{fig:examplegrowth} left). 
After a given period of $T$ simulation steps, the environment changes and the rightmost region gets the high profitability: $q_i=0.1, \forall i=1..7$ and $q_8=0.3$ (Fig.~\ref{fig:examplegrowth} right). 
In all the experiments reported here, $T=400$. 
The first switch occurs at simulation step $t=T$ and the experiments continue for another $T$ steps and then switch back to the initial environment at $t=2T$. The experiments end at $t=3T$.

\subsection{Topologies}
As illustrated in Fig.~\ref{fig:topologies}, the 
experiments are conducted for the following topologies:  
a) growable tree: a binary tree that initially starts with a root and two leaves and can adapt its shape by further growing or trimming branches; 
b) fixed tree: a static balanced binary tree with 4 leaves, where every leaf supports 2 regions of the environment;
c) line: 4 leaves where the neighboring leaves share a root;
d) circle: similar to a line except that the first and the last leaves share a root as well;
e) complete network: 4 leaves where every two leaves share a root;
f) all-to-root: 4 leaves all sharing a single root.

\subsection{Model parameters} 

All the experiments are initiated with 100 units of assets uniformly distributed among the service nodes. 
The growable topology is initiated with one root and two leaves. To grow new branches on a leaf node, the amount of assets in the node must exceed a threshold of $25$ units. Likewise, to trim a branch, a threshold of $20$ must be crossed (Cf. sec.\ref{sec:VMC}).
The management costs are $c=0$ for all the experiments.
The default values for  $\gamma_{\uparrow\!A}=\gamma_{\uparrow\!F}=\gamma_{\downarrow\!A}=\alpha = 1$. That is, there is no communication or operation delay, unless stated otherwise in the experiment.
The value of the competition factor $\beta$ is explicitly  stated for each experiment.


\begin{figure}[tp]
\centerline{
\includegraphics[width=0.4\textwidth, angle=0]{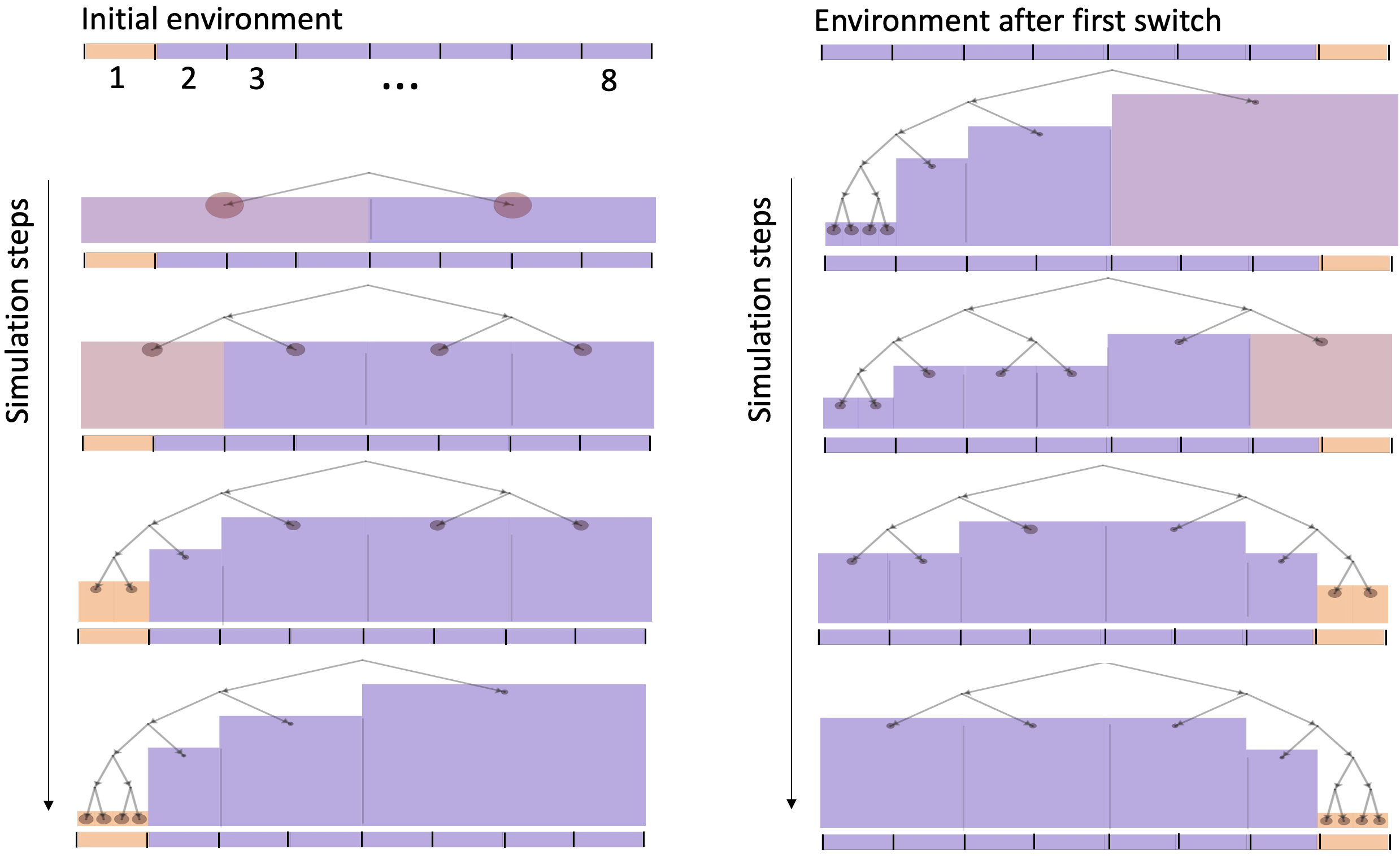}
}
\caption{A series of selected stages of growth. 
}
\label{fig:examplegrowth}
\end{figure}

\section{Experimental Results}

\begin{figure*}[htbp]
\centerline{
\includegraphics[width=0.6\textwidth, angle=0]{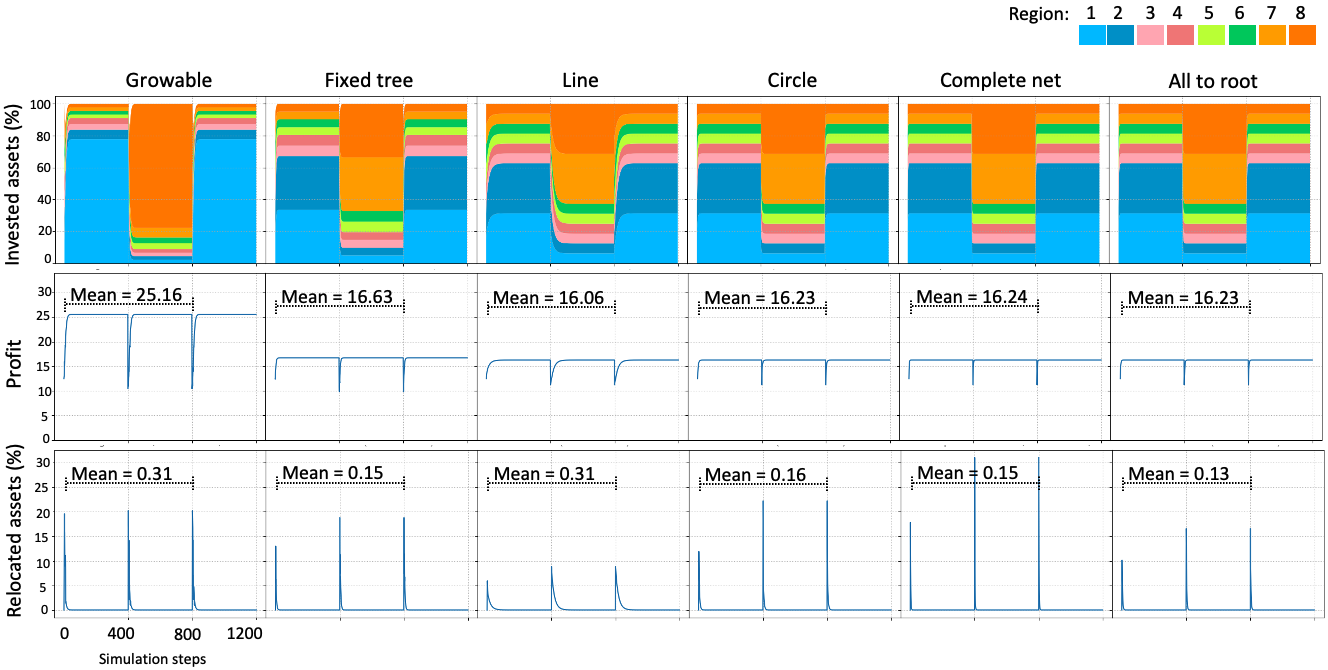}
}
\caption{competition factor = 0.7, no delay, no management cost.}
\label{fig:res1}
\end{figure*}

In the first set of experiments, a competition factor $\beta=0.7$ is used. 
This particular value of $\beta$ is selected based on preliminary experiments (not shown here) to represent a case where the behavioural dynamics are not drastically different for the various topologies.
The experiments start with the region 1 of the environment being the most profitable region. At simulation step 400, the environment changes and region 8 becomes the most profitable region. At simulation step 800, the environment switches back to the initial setup.
Fig.\ref{fig:res1} shows a) the asset investment per each region of the environment, b) the total profit produced within the system, and c) the percentage of relocated assets over time, for various topologies.
The results indicate a relatively slow response of line topology to the changes in the environment. 
The system with growable topology produces highest profit by effectively investing most of the assets into the best region.
Note that unlike growable topology that can grow deeper to specialise in service node regions, in all other topologies (static topologies) each service node is extended in a pair of neighboring regions -- indicated in the figure by different shades of the same color.

Fig.~\ref{fig:examplegrowth} shows a series of snapshots from various stages of structural adaptation of growable topology. Four snapshots are taken in the initial environment and 4 are taken after the first switch. As it is shown in the figure, the system starts with two leaves, each supporting 4 regions of the environment. The structure grows toward region 1 with the highest profitability while retracting from all the other regions. 
After the first switch, the structure starts retracting from the previously best region and grows toward the new best region 8.

\begin{figure*}[htbp]
\centerline{
\includegraphics[width=0.82\textwidth, angle=0]{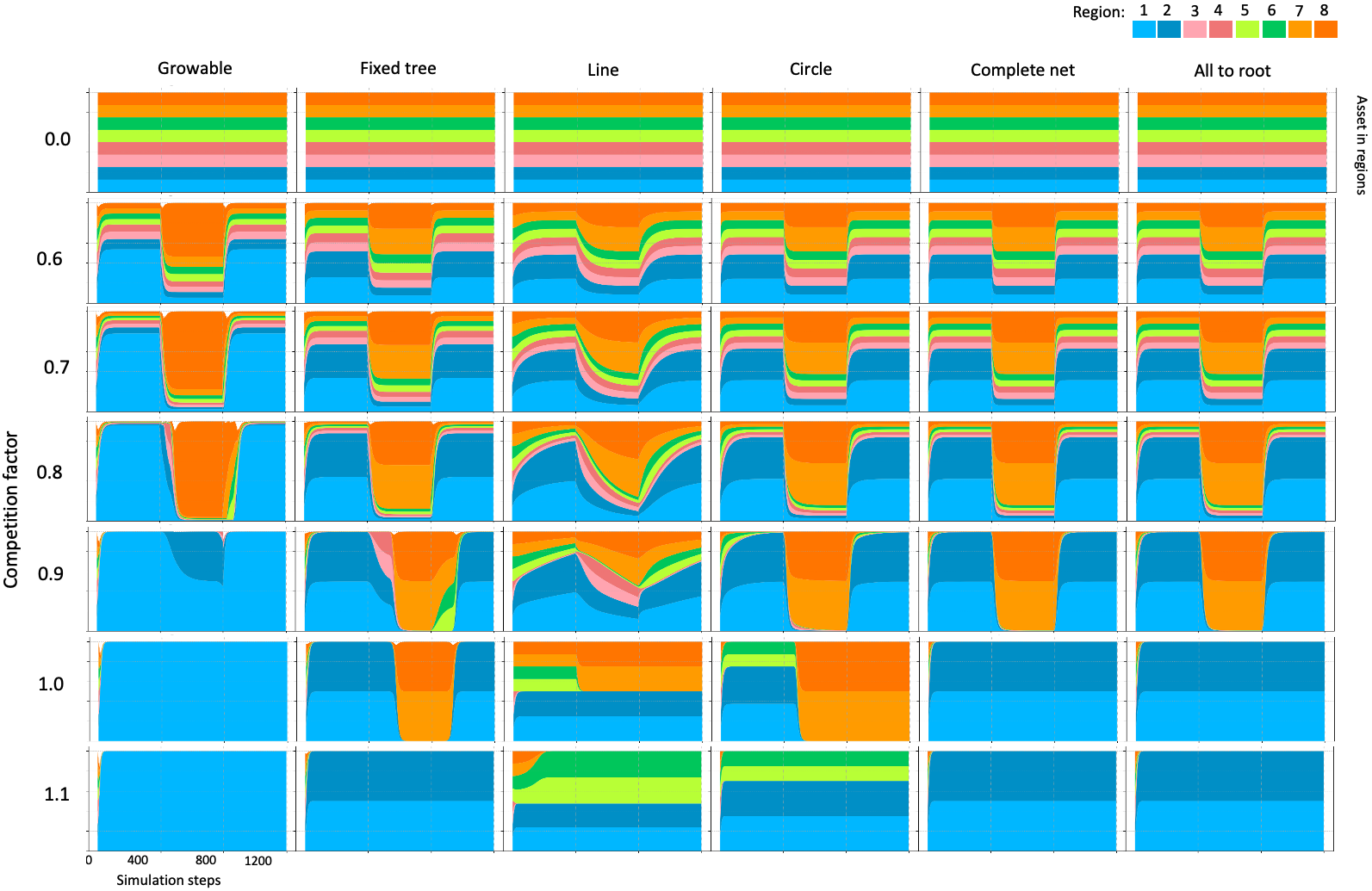}
}
\caption{
Comparison of asset investment dynamics in various regions of the environment for different competition factors $\beta$ and various topologies.
Results are shown for asset release ($\alpha=0.2$).}
\label{fig:cfCommparison}
\end{figure*}

The second set of experiments investigate the effects of various values for the competition factors $\beta$ on the dynamics of asset investment and profit production in systems with various topologies. 
Here, the release factor of assets ($\alpha$) is set to $0.2$ applying a delay in the asset relocation operation.
Fig.~\ref{fig:cfCommparison} shows the asset investment dynamics for various values of $\beta$ in different topologies.
Table.~\ref{tab:cfMeanProd} shows the average profit productions over the first 800 simulation steps of the same experiments. 
The highest and the lowest profit production levels (indicated in bold) belong to the growable topology at $\beta=0.8$ and the line topology at $\beta=1.1$ respectively.

The first row in Fig.~\ref{fig:cfCommparison} shows the results for $\beta=0$, leading to no competition between sibling branches irrespective of their profitability. That results in a uniform asset distribution in all the regions of the environment.
With larger $\beta$, the neighboring branches start to compete with each other based on their profitability. 
This leads to the attraction of assets to more profitable regions. 
As shown in the figure, when $\beta$ gets too large, the system loses its flexibility and adaptability to the environmental change. 
For different topologies, the effect kicks in at different $\beta$ values -- e.g. growable topology fails to adapt at $\beta=0.9$, while the fixed tree fails at $\beta=1.1$.  
For all values of $\beta$, the Line topology shows slower response to changes and a relatively lower profit production. 

Another interesting observation is in the fixed tree at $\beta=0.9$ (and with less clarity at $\beta=0.8$, also in growable topology at $\beta=0.8$ and $0.9$). 
In this case, after the environment switches, the speed of asset relocation to the newly profitable regions is very slow and almost invisible until a certain period. Meanwhile, assets move from the service nodes in the previously profitable regions to their siblings (e.g. from the blue regions to the pink ones), which is not beneficial in terms of profit production (profit production dynamic are not shown here).
After this first period and as soon as enough assets are relocated to the newly profitable regions, the relocation process gets a high speed until most of the assets are in the newly profitable regions.

\begin{table}[htbp]
\caption{Mean profit over 800 simulation steps, with $\alpha=0.2$}
\begin{center}
\begin{tabular}{|c|c|c|c|c|c|c|}
\hline
\textbf{$\beta$} & 
{{Growable}}& 
{{Fixed tree}}& 
{{Line}} & 
{{Circle}} & 
{{Complete}} & 
{{All-to-root}} \\
\hline
\cline{1-7}
0.0 & 12.5 & 12.5 & 12.5 & 12.5 & 12.5 & 12.5\\
\hline
0.6 & 20.2 & 15.0 & 14.5 & 14.8 & 14.8 & 14.8\\
\hline
0.7 & 24.5 & 16.5 & 15.3 & 16.0 & 16.1 & 16.0\\
\hline
0.8 & \textbf{26.2} & 18.3 & 15.9 & 17.9 & 18.1 & 18.0\\
\hline
0.9 & 19.6 & 17.6 & 14.9 & 19.0 & 19.4 & 19.4\\
\hline
1.0 & 19.6 & 17.6 & 14.9 & 17.5 & 14.9 & 14.9\\
\hline
1.1 & 19.6 & 14.9 & \textbf{12.3} & 13.4 & 14.9 & 14.9\\
\hline
\end{tabular}
\label{tab:cfMeanProd}
\end{center}
\end{table}

\section{Discussion}
\label{sec:discuss}

The most important insights we draw from the obtained results include the following.

   \textit{Growable versus static topologies: } 
   topologies that self-adapt via growth (`Growable') have higher potential to produce profits under changing conditions compared to static topologies. This is because growth allows to reallocate and re-specialise more assets in areas that become profitable. This comes at the cost of longer adaptation periods and lesser profits during these periods. In business organisations, reallocation and re-specialisation practices may also negatively impact worker experience.  
   In comparison, static topologies can reallocate assets but cannot re-specialise them. This limits their ability to adapt to (and profit from) new opportunities that were not predicted in their fixed structures. At the same time, they incur fewer asset re-allocations, because of lower generated profits; and shorter self-adaptation profit disturbance.
   
    \textit{Tree versus multi-root topologies: }
    in multi-root topologies (e.g. `line'), global control coordination occurs via indirect propagation of control information across roots (i.e. indirect horizontal coordination). This causes less effective asset re-allocation (i.e. going through ineffective intermediary states), inducing longer self-adaptation periods and impact on profits. This is similar to the impacts of communication delays in multi-scale structures (e.g. \cite{fgcs2021}), except propagating horizontally rather than vertically. 
    
    As these delay and perturbation effects are directly related to global control coordination, they tend to disappear when self-adaptation only requires local coordination. In the presented experiments, this was the case in the `circle' and `complete' topologies. 
    Here, the system parts concerned by the environmental changes were under the control of a single root, hence removing the need for coordination among several roots. For the same reasons, we note that multi-root topologies with complete meshing amongst roots behave similarly to single-scale tree topologies because all environment changes concern system parts that are  observed directly by one root, so no root coordination is needed. Yet, the former topology (`complete') induces more management and communication costs. 
    
    \textit{Competition level: }
    Some level of profit-oriented competition amongst system branches is necessary to drive self-adaptation-- otherwise assets are distributed equally across all environment areas, irrespective of their profitability and changes (Cf. Fig. \ref{fig:cfCommparison}, competition $= 0.0$). However, `extreme' competition hampers self-adaptation -- the actual degree depending on the competition degree and topological features. The main reason is that high competition favours high adaptation for short-term profits (e.g. over-fitting). Thus, when  profitability changes, there aren't enough assets to detect the new profit opportunities, as most assets have been allocated to the previously profitable areas. This incurs increasingly long adaptation delays, with inefficient intermediary adaptation states. e.g. in Fig. \ref{fig:cfCommparison} this can be observed for the `fixed tree' and `line' topologies, with competition factor $= 0.9$, via large pink areas, which indicate significant asset allocations to non-profitable areas (i.e. located in the middle of the environment bar, whereas the profitable areas are at the extremes). %
    Extreme competition ultimately leads to the inability to adapt, as the amount of assets allocated to the previously profitable area (that has become less profitable) is large enough to generate higher profits than the fewer assets allocated to the newly profitable area. This causes high inertia and blocks reactivity. 
   
   \textit{Single versus multi-scale of control: }
   Importantly, we only considered communication delays of one step across multi-scale topologies (just as for single-scale topologies). In most real systems this will not be the case as more scales will incur heavier control delays and potentially significant impacts on the self-adaptation outcomes (as we showed in previous work \cite{acsos2021}). Similarly, considering management costs at the decision nodes will weight heavier on multi-scale structures with more decision nodes (not shown in the results).
   
   In general, we can conclude that topologies and configurations that are more reactive (i.e. favouring more self-adaptation to changing environments) produce more profits on the short-term, while necessarily incurring  more adaptation costs. At the same time, too much reactivity (i.e. self-adaptation leading to over-fitting) can hamper further self-adaptation on the longer term. These findings are consistent with common observations from control and feedback systems, while providing specific insights into the asset distribution domain. The provided results offer useful intuitions into which kinds of topological characteristics and configurations favour more or less self-adaptation and the ensuing consequences.   
   
   \section{Conclusions}
   
   Within the asset-distribution domain, this paper studied the relation between multi-scale systems topology and the ensuing self-adaptative behaviour. We proposed a novel asset-distribution algorithm, generalising from previous work. We studied topologies with single or multiple roots; with growing capacities or static; and with multiple or single control scales. We also considered internal competition as an important concern across all topologies. Results led to the following key insights:
   1) topologies that allow asset re-specialisation (e.g. by `growing') are more adaptable to unforeseen opportunities and can provide higher profits than static topologies; 2) multi-root topologies incur higher delays than trees, due to extra inter-root communication delays (i.e. horizontal, same-scale communication) -- 
   this is similar to delays in multi-scale topologies (i.e. vertical inter-scale delays); 3) while some level of competition between system branches is needed for self-adaptation, too much competition actually hampers adaptation, by causing over-fitting and too much inertia. In all cases, more adaptation is achieved via more asset relocation and/or re-specialisation and hence incurs higher delays and disturbance. 
   These insights provide a basis for further studies, aiming to offer reusable guidance
   for selecting suitable system topologies in various dynamic environments.

\bibliographystyle{plain}
\bibliography{references.bib}

\begin{thebibliography}{10}

\bibitem{BALE2015}
Catherine~S.E. Bale, Liz Varga, and Timothy~J. Foxon.
\newblock Energy and complexity: New ways forward.
\newblock {\em Applied Energy}, 138:150--159, 2015.

\bibitem{bejan2012}
Adrian Bejan and J.~Peder Zane.
\newblock {\em Design in Nature: How the Constructal Law Governs Evolution in
  Biology, Physics, Technology, and Social Organizations}.
\newblock Doubleday, 2012.

\bibitem{energyDistrib2015}
Raffaele Carli and Mariagrazia Dotoli.
\newblock A decentralized resource allocation approach for sharing renewable
  energy among interconnected smart homes.
\newblock In {\em IEEE Conf. Dec. and Control (CDC)}, pages 5903--5908, 2015.

\bibitem{Crosland2010}
M.~W.~J. Crosland, S.~X. Ren, and J.~F.~A. Traniello.
\newblock Division of labour among workers in the termite, reticulitermes
  fukienensis (isoptera: Rhinotermitidae).
\newblock {\em Ethology}, 104:57--67, 2010.

\bibitem{cube-saso2012}
Bassem Debbabi, Ada Diaconescu, and Philippe Lalanda.
\newblock Controlling self-organising software applications with archetypes.
\newblock In {\em {IEEE} Int.Cnf.on Self-Adaptive and Self-Organizing Systems,
  {SASO}, Lyon, FR, Sept. 10-14, 2012}, pages 69--78.

\bibitem{diStefano2005}
Andrea Di~Stefano and Corrado Santoro.
\newblock A decentralized strategy for resource allocation.
\newblock volume~27, pages 295-- 300, 07 2005.

\bibitem{saso2019}
Ada Diaconescu, Louisa Jane~Di Felice, and Patricia Mellodge.
\newblock Multi-scale feedbacks for large-scale coordination in self-systems.
\newblock In {\em {IEEE} Intl.Cnf. on Self-Adaptive and Self-Organizing
  Systems, {SASO}, Umea, Sweden, June 16-20, 2019}, pages 137--142, 2019.

\bibitem{fgcs2021}
Ada Diaconescu, Louisa Jane~Di Felice, and Patricia Mellodge.
\newblock Exogenous coordination in multi-scale systems: How information flows
  and timing affect system properties.
\newblock {\em FGCS}, 114:403--426, 2021.

\bibitem{Huang92}
Z~Y Huang and G~E Robinson.
\newblock Honeybee colony integration: worker-worker interactions mediate
  hormonally regulated plasticity in division of labor.
\newblock {\em Proc Natl Acad Sci U S A}, 89(24):11726--9, 1992.

\bibitem{kounevSEAMS2011}
Nikolaus Huber, Fabian Brosig, and Samuel Kounev.
\newblock Model-based self-adaptive resource allocation in virtualized
  environments.
\newblock In {\em Intl.Symp.on Software Engineering for Adaptive and
  Self-Managing Systems (SEAMS)}, page 90–99, 2011.

\bibitem{JonesM03}
Chris Jones and Maja~J. Mataric.
\newblock Adaptive division of labor in large-scale minimalist multi-robot
  systems.
\newblock In {\em IEEE/RSJ Intl.Cnf. on Intelligent Robots and Systems (IROS)},
  pages 1969--1974, LA, 2003.

\bibitem{Julian99}
G.~E. Julian and Sara Cahan.
\newblock Undertaking specialization in the desert leaf-cutter ant acromyrmex
  versicolor.
\newblock {\em Anim Behav}, 58(2), 1999.

\bibitem{pautet2018}
Roberto Medina, Etienne Borde, and Laurent Pautet.
\newblock Scheduling multi-periodic mixed-criticality dags on multi-core
  architectures.
\newblock In {\em 2018 IEEE Real-Time Systems Symposium (RTSS)}, pages
  254--264, 2018.

\bibitem{acsos2021}
Patricia Mellodge, Ada Diaconescu, and Louisa~Jane Di~Felice.
\newblock Timing configurations affect the macro-properties of multi-scale
  feedback systems.
\newblock In {\em IEEE Intl. Cnf. ACSOS}, 2021.

\bibitem{barabasi2003}
Erzs{\'e}bet {Ravasz} and Albert-L{\'a}szl{\'o} {Barab{\'a}si}.
\newblock {Hierarchical organization in complex networks}.
\newblock {\em Physical Review E}, 67(2), Feb 2003.

\bibitem{arlesJAAMAS21}
Arles Rodriguez, Jonatan Gomez, and Ada Diaconescu.
\newblock Self-healing networks via self-organising mobile agents.
\newblock {\em Journal of Autonomous Agents and Multi-agent Systems (JAAMAS)},
  01 2021.

\bibitem{Seeley82}
Thomas~D. Seeley.
\newblock Adaptive significance of the age polyethism schedule in honeybee
  colonies.
\newblock {\em Behavioral Ecology and Sociobiology}, 11:287--293, 1982.

\bibitem{sardes2006}
Christophe Taton and et~al.
\newblock Self-sizing of clustered databases.
\newblock In {\em Intl. Symp. WoWMoM, Buffalo, New York, USA}, pages 506--512.
  {IEEE}, 2006.

\bibitem{Torres2012}
VO. Torres, TS. Montagna, J.~Raizer, and WF. Antonialli-Junior.
\newblock Division of labor in colonies of the eusocial wasp, mischocyttarus
  consimilis.
\newblock {\em Journal of insect science}, 12:21, 2012.

\bibitem{Yang2009}
Yongming Yang, Changjiu Zhou, and Yantao Tian.
\newblock Swarm robots task allocation based on response threshold model.
\newblock In {\em ICARA}, pages 171--176, Wellington, 2009. IEEE.

\bibitem{zahadat2019SASO}
Payam Zahadat.
\newblock Self-adaptation and self-healing behaviors via a dynamic distribution
  process.
\newblock In {\em 2019 IEEE 13th International Conference on Self-Adaptive and
  Self-Organizing Systems (SASO)}. IEEE, 2019.

\bibitem{zahadat2013ECALPSI}
Payam Zahadat, Karl Crailsheim, and Thomas Schmickl.
\newblock Social inhibition manages division of labour in artificial swarm
  systems.
\newblock In {\em EU Cnf. on Artificial Life (ECAL)}, pages 609--616. MIT
  Press, 2013.

\bibitem{ALife2020}
Payam Zahadat and Ada Diaconescu.
\newblock Reactive or stable: A plant-inspired approach for organisation
  morphogenesis.
\newblock In {\em The 2020 Conference on Artificial Life}, pages 614--622, 07
  2020.

\bibitem{zahadat2017gecco}
Payam Zahadat, Daniel~Nicolas Hofstadler, and Thomas Schmickl.
\newblock Vascular morphogenesis controller: A generative model for developing
  morphology of artificial structures.
\newblock In {\em Genetic and Evolutionary Computation Conference}, GECCO,
  pages 163--170. ACM, 2017.

\bibitem{zahadat2016adaptivebehavior}
Payam Zahadat and Thomas Schmickl.
\newblock Division of labor in a swarm of autonomous underwater robots by
  improved partitioning social inhibition.
\newblock {\em Adaptive Behavior}, 24(2):87--101, 2016.

\end{thebibliography}

\end{document}